\documentclass[prb,aps,twocolumn,amsmath,amssymb,floatfix,
superscriptaddress]{revtex4}

\usepackage[dvips]{graphics}
\usepackage{color}
\definecolor{dred}{rgb}{0,0,0.6}

\begin{document}

\title{Analytical study of nano-scale logical operations}

\author{Moumita Patra}

\affiliation{Physics and Applied Mathematics Unit, Indian Statistical
Institute, 203 Barrackpore Trunk Road, Kolkata-700 108, India}

\author{Santanu K. Maiti}

\email{santanu.maiti@isical.ac.in}

\affiliation{Physics and Applied Mathematics Unit, Indian Statistical
Institute, 203 Barrackpore Trunk Road, Kolkata-700 108, India}

\begin{abstract}

A complete analytical prescription is given to perform three basic (OR, AND, 
NOT) and two universal (NAND, NOR) logic gates at nano-scale level using 
simple tailor made geometries. Two different geometries, ring-like and 
chain-like, are taken into account where in each case the bridging conductor 
is coupled to a local atomic site through a dangling bond whose site energy 
can be controlled by means of external gate electrode. The main idea is that 
when injecting electron energy matches with site energy of local atomic
site transmission probability drops exactly to zero, whereas the junction
exhibits finite transmission for other energies. Utilizing this prescription 
we perform logical operations, and, we strongly believe that the proposed 
results can be verified in laboratory. Finally, we numerically compute 
two-terminal transmission probability considering general models and the 
numerical results matches exactly well with our analytical findings.

\end{abstract}

\maketitle

\section{Introduction}

Designing of logic gates has always been the subject of intense research
since these are considered as the basic building blocks of digital 
electronics. Among the widespread applications, logic gates carry the 
electronic information in a traditional computer system where around 
$100$ million gates are embedded to execute computational operations 
and these gates are made from field-effect transistors (FETs) and metal 
oxide semiconductor field effect transistors (MOSFETs).

The rapid progress of epitaxial and lithographic techniques has allowed us 
to fabricate 
phase-coherence-preserving samples where quantum interference effects could 
be observed~\cite{ref1}. Various novel electronic devices such as directional 
coupler, quantum stub transistor, electron Y-branch switch, and to name a few
have been proposed and analyzed~\cite{ref2,ref3,ref4}. These devices differ 
from ordinary electronic gadgets, as they are based on the quantum nature 
of electrons. More precisely, they rely on ballistic, non-phase-destroying
carrier transport, with the advantage of potentially gaining a speed increase 
of several orders of magnitude without dissipation. Exploiting the effect
of quantum interference~\cite{oj1,oj2,oj3,wal,int1,int2} one can construct 
logic gates using a device of atomic 
dimension. Following the pioneering work of de Silva and his group~\cite{ref5}
the idea of designing molecular logic gates~\cite{ref6,ref7,ref8,ref9,new1} 
has drawn much attention among researchers over last few decades assuming that
the logic gates at nano-scale level exhibit much better performance than 
the traditional logic gates made from FETs and MOSFETs. However, the main
limitation of constructing molecular logic gates is that they have much lower
gain as molecular systems have sufficiently low transconductance~\cite{ref10}.
Furthermore, after mounting hybrid molecular devices on a surface to construct 
complex circuits, surface tunneling leakage currents between the device and 
interconnects might still affect device performance~\cite{ref11}. 

Several other proposals have also been given for logical operations. For 
instance, photo-induced logic gates using DNA molecules and other 
nano structures~\cite{ref5a,ref11a}, though many controversial issues 
raised by researchers for its future capabilities in designing optical 
computers~\cite{ref11b}. For optical logic to be competitive, major 
breakthroughs in non-linear optical device technology should be required, 
or probably a change in the nature of computing itself. Other logical 
operations can also be implemented from quantum mechanical effects through 
quantum computing which usually diverges from Boolean design~\cite{ref11c}.

To find whether simple tailor made geometries are capable of designing
logic gates at nano-scale level, few years back one of the authors of us 
has suggested the possibilities of getting logical operations using 
simple quantum rings~\cite{ref12,sm1,sm2,sm3}. The key component for that 
model was the magnetic
\begin{figure}[ht]
{\centering \resizebox*{8cm}{5.5cm}{\includegraphics{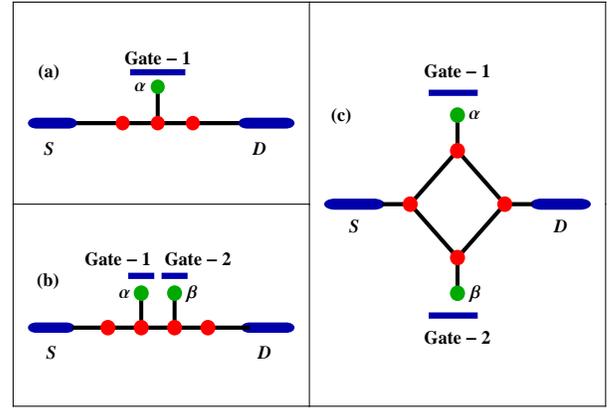}}\par}
\caption{(Color online). Schematic diagrams of three bridge setups
that we consider to explore logical operations where the bridging conductor
(chain- or ring-like) is coupled to local atomic site(s) ($\alpha$,
$\beta$) through dangling bond(s). The configuration (a) is used for NOT
operation (single logical operation), whereas each of the other two setups 
is used for two logical operations where (b) AND and NOR; and 
(c) OR and NAND. Gate-1 and Gate-2 are the gate electrodes through which
site energies of the sites $\alpha$ and $\beta$ are controlled.}
\label{scm}
\end{figure}
flux $\phi$ threaded by the ring, and it has to be fixed at half-flux quantum 
(i.e., $\phi=\phi_0/2$, where $\phi_0=ch/e$). Utilizing the concept of 
complete destructive interference under symmetric condition at $\phi=\phi_0/2$ 
all possible logic gates have been explored. But to confine this magnetic flux
in a nano-sized ring unrealistically large magnetic field is 
required~\cite{mag1,mag2,mag3}. This situation can be avoided by considering 
bigger rings. The crucial role of quantum interference effects (constructive 
and destructive) to achieve few logical operations using nano-rings has also 
been discussed in another work~\cite{ref12a} where the ring is coupled to 
three external leads, instead of two.

In the present work we intend to establish {\em how logical operations can
be achieved in simple two-terminal tailor made geometries without considering
any magnetic field, which of course yields an important step along this 
direction}. To do this we consider two different shaped conductors, 
chain-like and ring-like, and in each case the bridging conductor is coupled 
to local atomic site(s) through dangling bond(s) (for instance
see Fig.~\ref{scm} where general models of getting logical operations are
given. Detailed description of the models with more simplified versions 
are described in the appropriate places to analyze logical operations). 
The site energy of any such local atomic site can be altered with the help 
of an external gate electrode. Assigning two different site energies, 
associated with two different gate voltages, two states (ON and OFF) of 
an input signal are defined. Thus, for two-input logic gates we couple 
two such atoms with the parent conductor. The key idea is that whenever 
the injecting electron energy matches with site energy of the local 
atomic site complete suppression of electron transmission is obtained, while 
finite transmission is available for other cases. This phenomenon can be 
utilized to perform logical operations, and in the present work we explore 
all three basic logic gates (OR, AND, NOT) along with two universal gates 
i.e., NAND and NOR. A complete analytical prescription is given for these
logical operations, and finally, we numerically compute the results 
considering more general models those exactly matches with the analytical
findings. We strongly believe that the present proposal can be implemented 
through an experimental setup in laboratory.

The work is arranged as follows. In Sec. II we present theoretical 
prescription for a general model. The results are thoroughly analyzed in 
Sec. III, and finally, we conclude our findings in Sec. IV. 

\section{Theoretical Framework for a general bridge configuration: 
Transfer Matrix Method}

The logical response is described in terms of transmission probabilities
in a two-terminal setup. Finite (high) transmission corresponds to the 
`ON' state of the output, while the `OFF' state represents zero 
(or vanishingly small) transmission probability. Here, we present the
theoretical prescription based on transfer matrix (TM) 
method~\cite{tm1,tm2,tm3,tm4} for the calculation of two-terminal 
transmission probability considering a typical bridge setup, which can 
easily be utilized in other conducting junctions (those are given in 
Fig.~\ref{scm}) to explore different logical operations. 

\begin{figure}[ht]
{\centering \resizebox*{7.5cm}{2cm}{\includegraphics{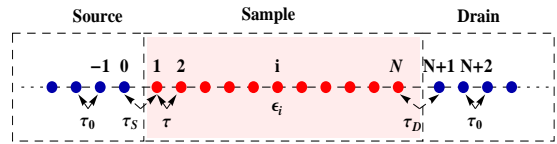}}\par}
\caption{(Color online). A quantum wire having $N$ atomic sites is coupled 
to two semi-infinite 1D electrodes, viz, source (S) and drain (D).}
\label{f1}
\end{figure}

Let us begin with Fig.~\ref{f1} where a one-dimensional (1D) chain is 
attached to source (S) and drain (D) electrodes through the coupling
parameters $\tau_S$ and $\tau_D$, respectively. For non-interacting case
one of the most important and usable models for studying electron
transport is the nearest-neighbor tight-binding (TB) model. In this
framework the TB Hamiltonian of the entire system is given by the sum
of three terms:~\cite{ref16} $H = H_{wire} + H_{elec} + H_{tun}$, where 
$H_{wire}$
represents the Hamiltonian of the bridging conductor sandwiched between
S and D, $H_{elec}$ and $H_{tun}$ correspond to the Hamiltonians of the
electrodes and chain-to-electrode coupling, respectively. These
Hamiltonians are written as:
\begin{eqnarray}
H_{wire} & = & \sum\limits_i \epsilon_i c_i^{\dagger} c_i + \sum\limits_i 
\tau \left(c_{i+1}^{\dagger} c_i + c_i^{\dagger} c_{i+1} \right)
\label{eq1}
\end{eqnarray}
\begin{eqnarray}
H_{elec} & = & H_S+H_D \nonumber \\
& = & \sum\limits_{n \le 0} \epsilon_0 a_n^{\dagger} a_n + 
\sum\limits_{n \le 0} \tau_0 \left(a_n^{\dagger} a_{n-1} + 
a_{n-1}^{\dagger} a_n \right) \nonumber \\
& & + \sum\limits_{n \ge N+1} \epsilon_0 b_n^{\dagger} b_n + 
\sum\limits_{n \ge N+1} \tau_0 \left(b_n^{\dagger} b_{n+1} + 
b_{n+1}^{\dagger} b_n \right)
\label{eq2}
\end{eqnarray}
and,
\begin{eqnarray}
H_{tun} &=& \tau_S \left(c_1^{\dagger}a_0 + a_0^{\dagger} c_1 \right)
+ \tau_D \left(c_N^{\dagger}b_{N+1} + b_{N+1}^{\dagger} c_N \right)
\label{eq3}
\end{eqnarray}
where $c_i^{\dagger}$ ($c_i$) corresponds to the creation (annihilation) 
operator, $\epsilon_i$ gives the on-site energy and $\tau$ represents the 
inter-atomic interaction between the neighboring atomic sites of the wire. 
$a_n^{\dagger}$ ($a_n$) and $b_n^{\dagger}$ ($b_n$) represent the creation 
(annihilation) operators in the source and drain electrodes, respectively, 
and they are parameterized by site energy $\epsilon_0$ and nearest-neighbor 
hopping integral $\tau_0$.

Now we discuss the method of calculating transmission probability for this 
setup using TM method. Let us start with the Schr\"{o}dinger equation 
\begin{equation}
H|\psi\rangle = E|\psi\rangle
\label{eq4}
\end{equation}
where the wave function $|\psi\rangle$ is expressed in terms of the Wannier
functions $|\psi_p\rangle$ of different sites as 
$|\psi\rangle = C_p|\psi_p\rangle$. $C_p$'s are the coefficients. In TM 
approach, the TM essentially connects the wave function of a particular site 
with its neighboring sites. For an arbitrary site $p$, its wave function 
$\psi_p$ (from now on we write $|\psi_p\rangle$ as $\psi_p$, for 
simplification) can be linked with the wave functions $\psi_{p+1}$ and 
$\psi_{p-1}$ of the neighboring sites through the transfer matrix 
(\mbox{\boldmath $P$}) as
\begin{equation}
\left(\begin{array}{cc}
    \psi_{i+1} \\ 
    \psi_i
\end{array}\right)
=\mbox{\boldmath $P$}\left(\begin{array}{cc}
    \psi_i \\ 
    \psi_{i-1}
\end{array}\right)
\label{eq5}
\end{equation} 
Thus for a $N$-site chain we can write the matrix equation as
\begin{equation}
\left(\begin{array}{cc}
    \psi_{N+2} \\ 
    \psi_{N+1}
\end{array}\right)
=\mbox{\boldmath $M$}\left(\begin{array}{cc}
    \psi_0 \\ 
    \psi_{-1}
\end{array}\right)
\label{eq6}
\end{equation}
where \mbox{\boldmath $M$}, representing the transfer matrix of the
full system, becomes
\begin{equation}
\mbox{\boldmath $M$}=\mbox{\boldmath $M_R$}\cdot\mbox{\boldmath $P_N$}\cdot
\mbox{\boldmath $P_{N-1}$}\cdot\cdot\cdot\cdot\mbox{\boldmath $P_{2}$}\cdot
\mbox{\boldmath $P_1$}\cdot\mbox{\boldmath $M_L$}
\label{eq7}
\end{equation}
where, \mbox{\boldmath $P_i$}'s are the transfer matrices for the sites 
labeled as $1, 2 ... N-1, N$, respectively, whereas,
\mbox{\boldmath $M_L$} and \mbox{\boldmath $M_R$} correspond to the
transfer matrices for the boundary sites at the left and right electrodes,
respectively. Solving Eq.~\ref{eq4} and doing some simple mathematical steps 
we get all the transfer matrices and they look like
\vskip 0.2cm
\noindent
$\mbox{\boldmath $M_L$}=\left(\begin{array}{cc}
    \frac{\tau_0}{\tau_S}e^{ik} & 0 \\ 
    0 & e^{ik}
\end{array}\right)$, $\mbox{\boldmath $M_R$}=\left(\begin{array}{cc}
    e^{ik} & 0 \\
    0 & \frac{\tau_D}{\tau_0}e^{ik}
\end{array}\right)$,\\
$\mbox{\boldmath $P_1$}=\left(\begin{array}{cc}
    \frac{E-\epsilon_1}{\tau} & -\frac{\tau_S}{\tau} \\
    1 & 0
\end{array}\right)$, $\mbox{\boldmath $P_N$}=\left(\begin{array}{cc}
    \frac{E-\epsilon_N}{\tau_{D}} & -\frac{\tau}{\tau_D} \\
    1 & 0
\end{array}\right)$, \\$\mbox{\boldmath $P_i$}=\left(\begin{array}{cc}
    \frac{E-\epsilon_i}{\tau} & -1 \\
    1 & 0
\end{array}\right)$, where $ 1<i<N $.
\vskip 0.2cm
\noindent
where $k$ is the wave vector. Assuming plane wave incidence with unit 
amplitude we can write the wave functions for the source and drain 
electrodes as
\vskip 0.2cm
\noindent
$\psi_n = e^{ikn} + r e^{-ikn}$, for $n \le 0$ \\
$~~~~~= t e^{ikn}$, for $n \ge N+1$
\vskip 0.2cm
\noindent
where $r$ and $t$ are the reflection and transmission coefficients.
Thus, Eq.~\ref{eq6} can be re-written as
\begin{equation}
\left(\begin{array}{cc}
    t e^{ik(N+2)} \\ 
    t e^{ik(N+1)}
\end{array}\right)
=\mbox{\boldmath $M$}\left(\begin{array}{cc}
    1+r \\
    e^{-ik} + r e^{ik}
\end{array}\right)
\label{eq9}
\end{equation}
Solving the above equation (Eq.~\ref{eq9}) we get the coefficient of 
transmission probability $t$ for each wave vector $k$, associated with 
incidence energy $E$, and eventually find the transmission probability
\begin{equation}
T(E)=|t|^2
\label{eq10}
\end{equation}
This is the general prescription for the calculation of transmission 
probability through a two-terminal conducting junction. Depending on the 
bridging system i.e., the conductor sandwiched between two electrodes we 
only modify the sub-Hamiltonian $H_{wire}$ keeping the rest unchanged.

Since all the logical operations described in the next section are analyzed
for very small sized conductors, the average spacing of successive energy 
levels are considerably large such that moderate temperatures have very 
minor impact on the proposed analysis. Therefore, we ignore the temperature
effect and set it to zero throughout the discussion. 

\section{Results and Discussion}

Based on the above theoretical prescription, worked out for a general model,
we analytically calculate transmission probabilities of different specific 
models required to achieve respective logical operations. Below we discuss 
them one by one.  
For the case of one-input logic gate (NOT gate) a single atomic site
having site energy $\epsilon_{\alpha}$ is coupled with parent lattice via
a dangling bond. This coupling is described by the parameter $\tau_{\lambda}$.
Whereas two such atoms are considered in the case of two-input logic gates.
Their site energies are defined as $\epsilon_{\alpha}$ and $\epsilon_{\beta}$,
and they are coupled with the identical strength $\tau_{\lambda}$, like 
one-input gate. 
The site energies $\epsilon_{\alpha}$ and $\epsilon_{\beta}$, treated as 
the inputs, are controlled by suitable gate electrodes. When no voltage is 
applied in the electrodes $\epsilon_{\alpha}=\epsilon_{\beta}=0$ and
it is called as the OFF state of the two inputs, whereas in presence of 
finite gate voltage $\epsilon_{\alpha}=\epsilon_{\beta}=1$ which is 
defined as ON state of the inputs. For simplification of analytical 
calculations we fix all the inter-atomic interactions ($\tau, \tau_0, \tau_S, 
\tau_D$, and $\tau_{\lambda}$) at $1\,$eV and choose site energies of all the 
atomic sites apart from local atomic site(s) to zero, though one can choose
any other set of parameter values. Only the thing is that the expressions will
be quite longer and difficult to read, but the physics will be exactly same. 
Thus, in the OFF state condition ($\epsilon_{\alpha}=\epsilon_{\beta}=0$), 
site energies of the local atomic sites become identical to the parent 
lattice sites.

\begin{figure}[ht]
{\centering \resizebox*{7.5cm}{5.5cm}{\includegraphics{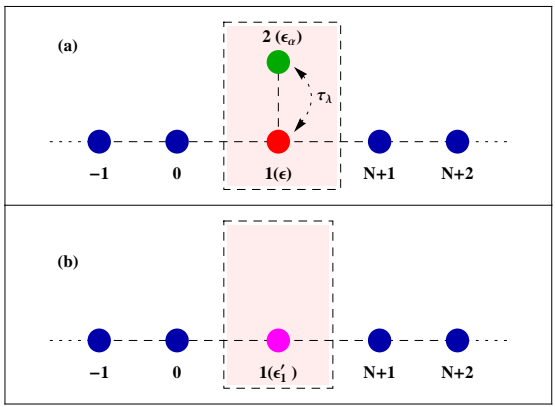}}\par}
\caption{(Color online). (a) Setup for NOT gate. The pink shaded region
describes the bridging conductor which contains one parent lattice site,
numbered as $1$, and it is directly coupled to a local atomic site, labeled
as $2$, having site energy $\epsilon_{\alpha}$. (b) After re-normalizing 
the configuration given in (a) where the effect of site $2$ is incorporated 
into the site $1$ to get a $1$D lattice. The change of color of site $1$ in 
(b) compared to (a) represents the site gets modified after re-normalization.}
\label{f2}
\end{figure}

\vskip 0.2cm
\noindent
\underline{\textbf{NOT gate}}: The bridge setup for NOT gate operation is
schematically shown in Fig.~\ref{f2}(a) in which the shaded region
represents the bridging conductor where a parent site (labeled as $1$)
is coupled to a local site (site number 2) via a dangling bond. The parent
site is again directly coupled to the measuring electrodes. To find
transmission probability for this setup first we write difference equations 
for these two sites using the Schr\"odinger equation (Eq.~\ref{eq4}) and 
they are
\begin{equation}
E\psi_1=\psi_0 + \psi_{N+1} + \psi_2
\label{eq11}
\end{equation}
\begin{equation}
\left(E-\epsilon_{\alpha}\right)\psi_2
= \psi_1
\label{eq12}
\end{equation}
These equations look very simple as we choose all the inter-atomic 
interactions at $1\,$eV and fix site energies of the electrodes and parent 
lattice sites to zero. Now substituting $\psi_2$ we get the renormalized 
difference equation for site $1$  as 
\begin{equation}
\left(E-\epsilon_1'\right)\psi_1=\psi_0 + \psi_{N+1}
\label{eq13}
\end{equation}
where the renormalized site energy of site $1$ becomes
\begin{equation}
\epsilon_1' = \frac{1}{E - \epsilon_{\alpha}}
\label{eq14}
\end{equation}
Thus the setup given in Fig.~\ref{f2}(a) maps exactly to the effective
$1$D configuration, (Fig.~\ref{f2}(b)), and therefore, we can utilize
the TM method to calculate transmission function $T$. Here we solve 
Eq.~\ref{eq9} considering $\mbox{\boldmath $M$}=\mbox{\boldmath $M_R$}\cdot
\mbox{\boldmath $P_1$}\cdot\mbox{\boldmath $M_L$}$ with
$\mbox{\boldmath $P_1$} = \left(\begin{array}{cc}
    E-\epsilon_1' & -1 \\
    1 & 0
\end{array}\right)$.
Doing the necessary steps we eventually reach to the expression of
transmission probability as
\begin{equation}
T=\frac{1}{1 + \frac{1}{\left(4 -E^2\right)
\left(E-\epsilon_{\alpha}\right)^2}}
\label{eq15}
\end{equation}
At a first glance it seems that the energy $E$ is 
dimensionless since transmission probability does not have any dimension. 
But this is not the case at all. We reach to this relation Eq.~\ref{eq15} 
only due to the fact that we set all the site
energies apart from local atomic site(s) ($\alpha$ and/or $\beta$) to 
zero, and fix the inter-atomic hopping integrals at $1\,$eV.  From the 
expression Eq.~\ref{eq15} we can clearly explain the NOT gate operation 
setting the injecting electron energy $E$ at $1\,$eV. When the input is 
high i.e.,
\begin{table}[ht]
\begin{center}
\caption{Implementation of NOT behavior setting $E=1\,$eV.}
\label{NOT}
~\\
\begin{tabular}{c c}
   \hline
  ~~Input~~ & ~~Output~~ \\
  \hline
  0 & 0.75 \\
  1 & 0 \\
 \hline
\end{tabular}
\end{center}
\end{table}
$\epsilon_{\alpha}=1\,$eV, transmission probability drops exactly to zero
(OFF state), while  high transmission viz $T=0.75$ (ON state) is obtained
when the input is low ($\epsilon_{\alpha}=0$). The response is summarized
in Table~\ref{NOT}. 

In order to check the robustness of our analysis for any such general model
\begin{figure}[ht]
{\centering \resizebox*{7.25cm}{4.75cm}{\includegraphics{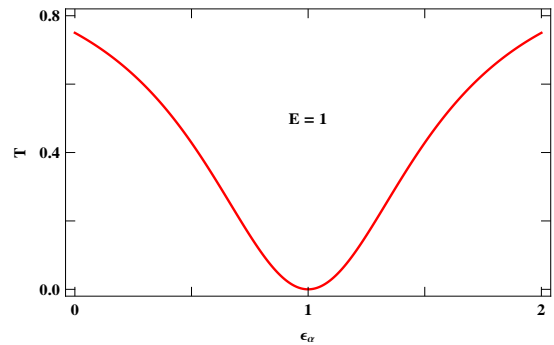}}\par}
\caption{(Color online). Two-terminal transmission probability as a function
of $\epsilon_{\alpha}$ for a general setup as given in Fig.~\ref{scm}(a)
considering $N=40$, $N_{\alpha}=20$ and $E=1\,$eV. $N_{\alpha}$ represents 
the location of parent lattice site where the local atomic site is coupled.}
\label{f3}
\end{figure}
in Fig~\ref{f3} we plot $T$-$\epsilon_{\alpha}$ characteristics which
we calculate numerically for a $40$-site chain considering $N_{\alpha}=20$ 
($N_{\alpha}$ represents the location of the parent atomic site with which 
the local site is coupled through the dangling bond). Figure~\ref{f3} 
clearly describes the NOT gate operation i.e., $T=0$ for 
$\epsilon_{\alpha}=1\,$eV and $T\simeq0.75$ when $\epsilon_{\alpha}=0$. Thus 
the numerical results exactly corroborate our analytical findings. In this
context it is important to note that one can choose any $N_{\alpha}$ as 
the results are independent of this position.

\begin{figure}[ht]
{\centering \resizebox*{7.5cm}{5.5cm}{\includegraphics{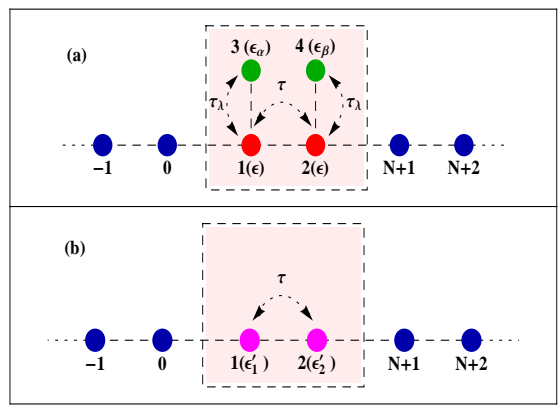}}\par}
\caption{(Color online). (a) Conducting junction for AND and NOR operations.
The bridging conductor contains two parent lattice sites ($1$ and $2$) those
are coupled to two local atomic sites ($3$ and $4$) having site energies
$\epsilon_{\alpha}$ and $\epsilon_{\beta}$. (b) Renormalized version
of configuration (a).}
\label{f4}
\end{figure}

\vskip 0.2cm
\noindent
\underline{\textbf{AND and NOR gates}}: To achieve AND and NOR gates,
the setup is slightly modified than the previous one i.e., the
configuration used for NOT gate. Using the bridge configuration, given
in Fig.~\ref{f4}(a), we can get both AND and NOR operations
{\em which seems very interesting as only one setup provides two different 
logical operations}. Here two atomic sites, labeled as $3$ and $4$,
having energies $\epsilon_{\alpha}$ and $\epsilon_{\beta}$ are connected
to the parent sites $1$ and $2$ respectively. In one step re-normalization
the system maps exactly to the $1$D lattice (see Fig.~\ref{f4}(b)) where
site energies of atomic sites $1$ and $2$ get modified as
\begin{equation}
\epsilon_1' = \frac{1}{E - \epsilon_{\alpha}} ~~~\mbox{and}~~~
\epsilon_2' = \frac{1}{E - \epsilon_{\beta}}
\label{eq16}
\end{equation}
Under this situation the full transfer matrix $\mbox{\boldmath $M$}$ becomes
$$\mbox{\boldmath $M$}=\mbox{\boldmath $M_R$}
\cdot\mbox{\boldmath $P_2$}\cdot\mbox{\boldmath $P_1$}
\cdot\mbox{\boldmath $M_L$}$$
with
\vskip 0.2cm
\noindent
$\mbox{\boldmath $P_1$} =
\left(\begin{array}{cc}
    E-\epsilon_1' & -1 \\
    1 & 0
\end{array}\right)$ and $\mbox{\boldmath $P_2$} =
\left(\begin{array}{cc}
    E-\epsilon_2' & -1 \\
    1 & 0
\end{array}\right)$.
\vskip 0.2cm
\noindent
Finally solving Eq.~\ref{eq9} we get the following expression of
transmission probability
\begin{widetext}
\begin{equation}
T=\frac{(4 -E^2)(E-\epsilon_{\alpha})^2(E-\epsilon_{\beta})^2}
{(\epsilon_{\alpha}+\epsilon_{\beta}-3\cos k)^2+\{(1+ 2\epsilon_{\alpha}
\epsilon_{\beta})\sin k - 2(\epsilon_{\alpha} + \epsilon_{\beta})\sin 2k +
2\sin3k\}^2}
\label{eq17}
\end{equation}
\end{widetext}
From this expression we can explain two logical operations (AND and NOR)
setting the energy $E$ at two different values. For $E=0$, we get AND
operation and the corresponding truth table is given in Table.~\ref{AND}.
\begin{table}[ht]
\begin{center}
\caption{Truth table for AND gate at $E=0$.}
\label{AND}
~\\
\begin{tabular}{c c c}
   \hline
  ~~Input-I~~ & ~~Input-II~~ & ~~Output~~ \\
  \hline
  0 & 0 & 0 \\
  0 & 1 & 0 \\
  1 & 0 & 0 \\
  1 & 1 & 0.75 \\
 \hline
\end{tabular}
\end{center}
\end{table}
\begin{table}[ht]
\begin{center}
\caption{Truth table for NOR gate at $E=1\,$eV.}
\label{NOR}
~\\
\begin{tabular}{c c c}
   \hline
  ~~Input-I~~ & ~~Input-II~~ & ~~Output~~ \\
  \hline
  0 & 0 & 1 \\
  0 & 1 & 0 \\
  1 & 0 & 0 \\
  1 & 1 & 0 \\
 \hline
\end{tabular}
\end{center}
\end{table}
Whereas for $E=1\,$eV, the NOR operation is obtained and its truth table is
shown in Table.~\ref{NOR}.

\begin{figure}[ht]
{\centering \resizebox*{9.5cm}{9cm}{\includegraphics{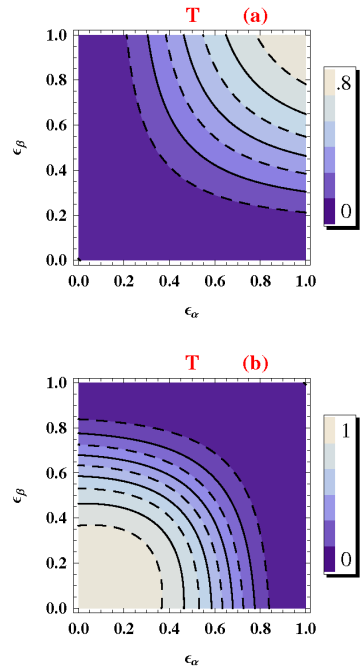}}\par}
\caption{(Color online). Density plot of two-terminal transmission
probability as functions of $\epsilon_{\alpha}$ and $\epsilon_{\beta}$
to analyze AND (shown in (a)) and NOR (shown in (b)) operations, for
a general model given in Fig.~\ref{scm}(b), taking $N=60$,
$N_{\alpha}=40$ and $N_{\beta}=45$, ($N_{\beta}$, similar to $N_{\alpha}$,
represents the location of parent site coupled to the local site having
site energy $\epsilon_{\beta}$). In (a) we set $E=0$ while in (b)
we fix $E=1\,$eV.}
\label{f5}
\end{figure}

In Fig.~\ref{f5} we present the density plot of two-terminal transmission
probability $T$, computed numerically as functions of $\epsilon_{\alpha}$ 
and $\epsilon_{\beta}$ for two different energies
($E=0$ and $1\,$eV), for a general model, considering $N=60$, $N_{\alpha}=40$
and $N_{\beta}=45$. From the spectra it is shown that only when both the two
inputs are high the output is high at $E=0$ yielding AND operation
(Fig.~\ref{f5}(a)). On the other hand, the same bridge setup exhibits high
output only when both the inputs are low under the condition
$E=1\,$eV resulting the NOR operation (see (Fig.~\ref{f5}(b)). The
numerical results completely match with analytical observations.

\begin{figure}[ht]
{\centering \resizebox*{8.5cm}{11cm}{\includegraphics{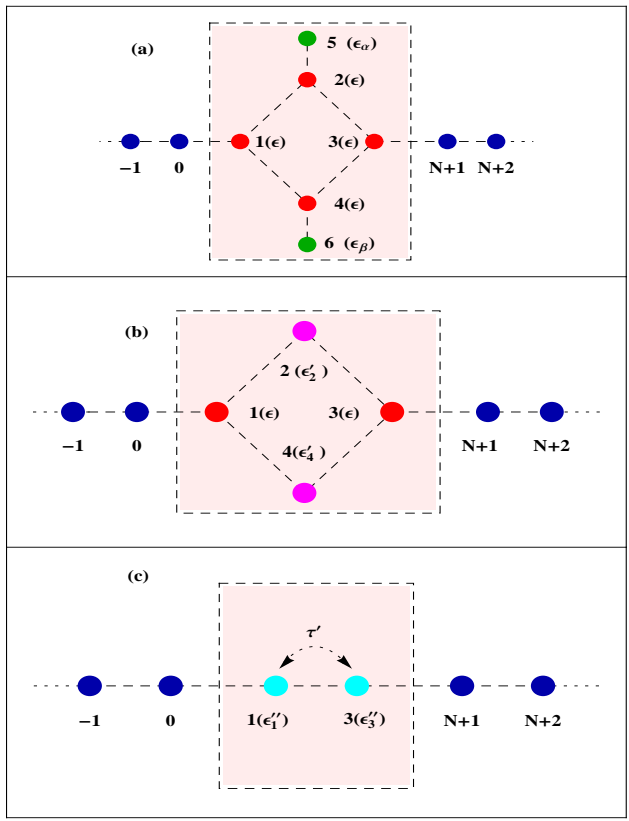}}\par}
\caption{(Color online). (a) Bridge configuration for OR and NAND operations
where the pink shaded region contains a diamond shaped conductor which is
coupled to two local atomic sites ($5$ and $6$). (b) After one step
re-normalization an effective diamond is formed by incorporating the
effects of sites $5$ and $6$ into $2$ and $4$, respectively. (c) The diamond
becomes an effective bond in second step re-normalization. Thus the
final system becomes a linear chain.}
\label{f6}
\end{figure}

\vskip 0.2cm
\noindent
\underline{\textbf{OR and NAND gates}}: Finally, we focus on another setup,
given in Fig.~\ref{f6}(a), that is used to exhibit OR and NAND operations.
Here a diamond like interferometric geometry is taken into account instead
of a chain-like which is connected to two atomic sites having site energies
 $\epsilon_{\alpha}$ and $\epsilon_{\beta}$  via two dangling bonds. Here
two step re-normalizations are required to get an effective $1$D lattice.
In the first step, the effects of $5$th and $6$th atomic sites are
incorporated into site numbers $2$ and $4$, respectively, to get a regular
diamond shaped conductor (see Fig.~\ref{f6}(b)). Under this operation the site
energies of these two sites ($2$ and $4$) become
\begin{equation}
\epsilon_2' = \frac{1}{E - \epsilon_{\alpha}}~~~\mbox{and}~~~
\epsilon_4' = \frac{1}{E - \epsilon_{\beta}}
\label{eq18}
\end{equation}
In the second step re-normalization, diamond shaped conductor moves to a
linear one (see Fig.~\ref{f6}(c)) where effectively a bond is formed
among the sites $1$ and $3$. Both their site energies and inter-atomic
hopping integral get modified as,
\begin{equation}
\epsilon_1'' = \epsilon_2'' = \tau'=\frac{E - \epsilon_{\alpha}}
{E (E - \epsilon_{\alpha})-1} + \frac{E - \epsilon_{\beta}}
{E (E - \epsilon_{\beta})-1}
\label{eq19}
\end{equation}
Once we get this effective linear geometry we can employ $TM$ method, like
previous cases, and the transmission probability becomes,
\begin{equation}
T = \frac{x^2(4-E^2)}{1 - 2 x E + 4 x^2}
\label{eq20}
\end{equation}
where $x = \frac{E - \epsilon_{\alpha}}
{E (E - \epsilon_{\alpha})-1} + \frac{E - \epsilon_{\beta}}
{E (E - \epsilon_{\beta})-1}$.
The above expression clearly describes other two logical operations, like
\begin{figure}[ht]
{\centering \resizebox*{9.5cm}{8.5cm}{\includegraphics{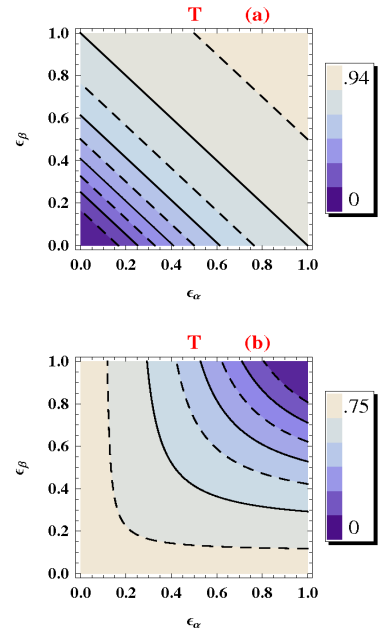}}\par}
\caption{(Color online). Density plot of two-terminal transmission 
probability as functions of $\epsilon_{\alpha}$ and $\epsilon_{\beta}$ 
to characterize OR (shown in (a)) and NAND (shown in (b)) operations, 
for the model given in Fig.~\ref{scm}(c). In (a) we choose $E=0$ while 
in (b) we take $E=1\,$eV.}
\label{f7}
\end{figure}
previous setup, by setting injecting electron energy at two distinct values. 
For $E=0$ we get OR operations, whereas for $E=1\,$eV NAND operation is
\begin{table}[ht]
\begin{center}
\caption{Truth table for OR gate at $E=0$.}
\label{OR}
~\\
\begin{tabular}{c c c}
   \hline
  ~~Input-I~~ & ~~Input-II~~ & ~~Output~~ \\
  \hline
  0 & 0 & 0 \\
  0 & 1 & 0.8 \\
  1 & 0 & 0.8 \\
  1 & 1 & 0.94 \\
 \hline
\end{tabular}
\end{center}
\end{table}
is observed. And their corresponding truth tables are given in
Table.~\ref{OR} and Table.~\ref{NAND}, respectively.
\begin{table}[ht]
\begin{center}
\caption{Truth table for NAND gate at $E=1\,$eV.}
\label{NAND}
~\\
\begin{tabular}{c c c}
   \hline
  ~~Input-I~~ & ~~Input-II~~ & ~~Output~~ \\
  \hline
  0 & 0 & 0.75 \\
  0 & 1 & 0.75 \\
  1 & 0 & 0.75 \\
  1 & 1 & 0 \\
 \hline
\end{tabular}
\end{center}
\end{table}
At the end, in
Fig.~\ref{f7} we present the density plot of two-terminal transmission
probability by varying $\epsilon_{\alpha}$ and $\epsilon_{\beta}$ to
check the sensitivity of logical operations on these quantities. We find
that the logical operations are stable for a wide range of
parameter values, and thus, can be tested in laboratory.

\section{Closing Remarks}

In the present work we intend to establish how logical operations can be 
performed using simple tailor made geometries. Comparing all the 
propositions we can argue that our proposed model is the most suitable one, 
particularly due to its simplicity, to design logic gates at nano-scale 
level. A complete analytical prescription is given to understand three basic 
(OR, AND, NOT) and two universal (NAND and NOR) logic gates. Two geometrical 
shapes of the bridging conductor, chain-like and ring-like, are taken into 
account where in each case the conductor is coupled to a local atomic site 
or two such sites depending on one input or two input logic gates. By using 
gate electrode site energy of the local site is controlled which determines 
the OFF or ON state of an input signal. The key idea is that whenever the 
site energy of a local atomic site matches with the injecting electron energy, 
transmission probability drops exactly to zero. This phenomenon is utilized 
to design the logical operations. In this paper we propose five logical 
operations and unable to establish other two operations viz, XOR and XNOR 
gates using such simple setups. Hopefully we will do that in our forthcoming 
work.

Finally, we would like to state that all these results are valid for a 
reasonable range of parameter values which we confirm through our
extensive numerical analysis and here we present some of them considering 
more general models with higher number of lattice sites. Thus, we strongly 
believe that the present proposal can be easily tested in laboratory.

\section{Acknowledgment}

MP would like to thank University Grants Commission, India 
(F. 2-10/2012(SA-I)) for her research fellowship.

\end{document}